\documentclass[12pt]{article}
\input{epsf.tex}
\usepackage{cite}
\begin{document}
\large
\begin{center}
{\bf ELRADGEN: Monte Carlo generator for radiative\\
events in elastic electron--proton scattering}

\vspace{3mm}
A.V. Afanasev

\vspace{3mm}
{\it Jefferson Lab, Newport News, VA 23606, USA}

\vspace{3mm}
I. Akushevich

\vspace{3mm}
{\it Duke University, Durham, NC 27708, USA}

\vspace{3mm}
A. Ilyichev

\vspace{3mm}
{\it Center of Particle and High Energy Physics,
220040 Minsk, Belarus}

\vspace{3mm}
B. Niczyporuk

\vspace{3mm}
{\it Jefferson Lab, Newport News, VA 23606, USA}
\end{center}
\normalsize
\begin{abstract}
We discuss the theoretical approach and practical algorithms for simulation
of radiative events in elastic $ep$--scattering.
A new Monte Carlo generator for real photon emission
events in the process of elastic
electron--proton scattering is presented.
We perform a few consistency checks and present numerical results.
\end{abstract}

\section{Introduction}

Radiative effects accompany any processes with charged light particle
scattering. These effects are exactly calculable in QED and they should
be incorporated as an essential step in the data analysis.
For elastic processes, which are the subject of the present report,
our starting point for the radiative effect
calculation is the technique proposed by Bardin and Shumeiko \cite{BSh}
and later developed for polarized particles as well as for specific
elastic and exclusive processes in papers
\cite{Ak,AAM,AAIM,AABJ}. For other approaches, please see Refs. \cite{MoTsai,MaxTj,ESF}.
In some cases the radiative effects can
be calculated analytically. However, often the analytical integration over the
photonic phase space is complicated because of specifics
of detector geometry or is not desirable because of experimental setup (for example,
detector resolution and kinematic cuts). In these situations an approach based
on Monte Carlo simulation of the radiative effects is the most adequate.
Monte Carlo generator RADGEN \cite{RADGEN} for deep inelastic processes was
developed on the basis of corresponding analytical code
POLRAD \cite{POLRAD}  and successfully applied by several experimental
collaborations analyzing polarized and unpolarized observables
in deep inelastic electron--nucleon scattering.

The reason for careful implementation of the radiative correction
procedure in elastic $ep$--scattering is twofold. First, elastic
$ep$--processes are commonly used for detector calibration,
and JLab CLAS detector is one of the examples.
Second, a new generation of experiments on elastic $ep$-scattering provide
data with very high accuracy (see, for instance, Refs. \cite{JLab1,JLab2}), and set new precision
requirements to data analysis and, in particular, to the radiative
correction procedure. In this report we present a new Monte Carlo generator for
elastic unpolarized $ep$-scattering events taking into account QED radiative
effects.
The main requirement for construction of the generator is
appearance of
simulated events (Born and radiative) as well as all event
distributions with respect to the kinematic variables in
accordance with their
contribution to the total cross section.

The contribution to the total cross section from the Born process
\begin{equation}
e(k_1)+p(p_1)\longrightarrow e'(k_2)+p'(p_2)
\end{equation}
is described by the transfer momentum squared $Q^2=-(k_1-k_2)^2$
and the azimuthal angle  $\phi$.
The cross section does not depend on the angle $\phi$ up to acceptance
effects. To describe the phase space
of the radiative process
\begin{equation}
e(k_1)+p(p_1)\longrightarrow e'(k_2)+p'(p_2)+\gamma (k),
\end{equation}
three new kinematic variable have to be considered.
We choose them as the proton transfer momentum squared
$t=-(k_1-k_2-k)^2$,
the inelasticity $v=(p_2+k)^2-M^2$ and the azimuthal angle $\phi_k$
between the planes ${\bf(q,k_1)}$ and  ${\bf
(k_1,k_2)}$.
This set of the variables defines the four--vectors of all final
particles in any frame.
Usually Lab system is used with OZ axis along the beam direction and plane OZX
parallel to the ground  level.

Thus, the strategy for simulation of one event can be defined as follows.
\begin{itemize}
\item For the fixed initial energy and $Q^2$ the non--radiative
and radiative parts of the total cross section are calculated.
Note, that together with Born subprocess, the contributions of
loop effects and soft photon radiation are also included into the
non--radiative
part.
\item The channel of scattering is simulated for this event
in accordance with partial contributions of these two positive parts
into the total cross section.
\item For any process $\phi$ is simulated uniformly from $0$ to
$2\pi$.
\item For the radiative event the kinematic variables $t$, $v$ and
$\phi_k$
are simulated in accordance with their calculated distribution (see
discussion below).
\item The four--momenta of all final particles in the required
system form are calculated.
\item If the initial $Q^2$ has not a fixed value but is instead simulated
according to some probability distribution (for example,
the Born cross section) then the cross sections have to be stored
for reweighting.
The  $Q^2$ distribution is simulated over the Born cross section,
and realistic observed $Q^2$ distribution is calculated as sum of
weights, they are ratios of the total and Born cross sections.
\end{itemize}

We note that separation of the bremsstrahlung process into the radiative
and nonradiative parts necessarily requires to introduce an additional
separating parameter. Usually this quantity is associated with photon
energy resolution in the detector. In our program the minimal
inelasticity $v_{min}$ is chosen  as this parameter.

\section{Non-radiated and radiated contributions}

The radiative events are generated in accordance with their
contribution to the observed total cross section, that can
be separated into non--radiative
and radiative parts:
\begin{equation}
\sigma _{obs} =
\sigma ^{non-rad}(v_{min}) +
\sigma ^{rad}(v_{min}).
\end{equation}
Here and later we define $\sigma \equiv d\sigma /dQ^2 d\phi$.
The first part includes the Born
cross section $\sigma _{0}$ (Fig.1 (a)), loop effects
(Fig.1 (b,c)) contributions as well as soft photon radiation, which is
restricted by
the inelasticity value $v<v_{min}$:
\begin{figure}[t]
\vspace*{50mm}
\hspace*{-16mm}
\unitlength 1mm
\begin{tabular}{ccccc}
\begin{picture}(20,20)
\put(0,0){
\epsfxsize=6cm
\epsfysize=10cm
\epsfbox{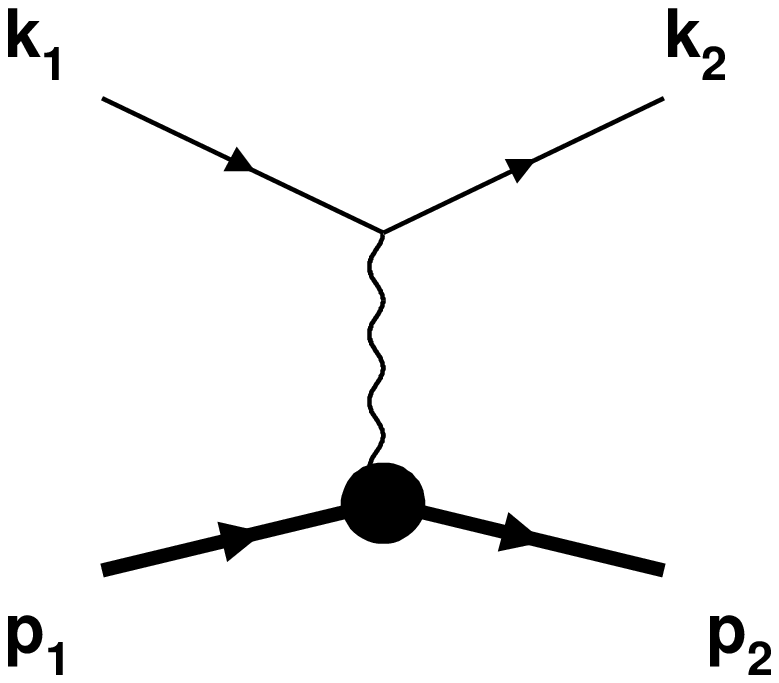}
\put(-36,36){\mbox{(a)}}
}
\end{picture}
&
\begin{picture}(20,20)
\put(10,0){
\epsfxsize=6cm
\epsfysize=10cm
\epsfbox{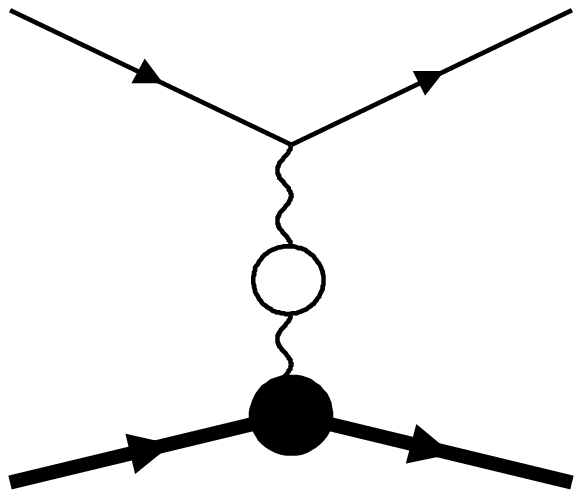}
\put(-36,36){\mbox{(b)}}
}
\end{picture}
&
\begin{picture}(20,20)
\put(20,0){
\epsfxsize=6cm
\epsfysize=10cm
\epsfbox{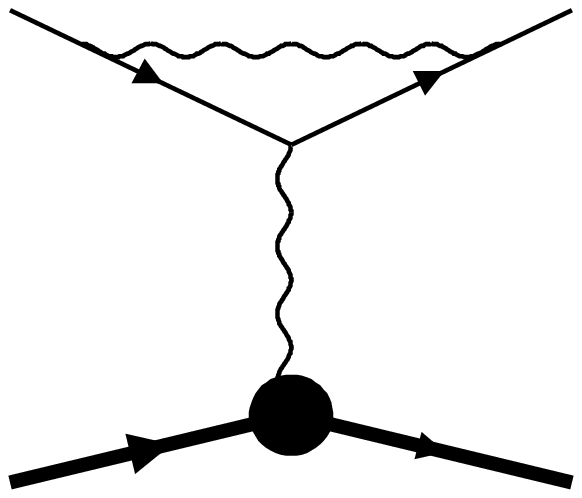}
\put(-36,36){\mbox{(c)}}
}
\end{picture}
&
\begin{picture}(20,20)
\put(30,0){
\epsfxsize=6cm
\epsfysize=10cm
\epsfbox{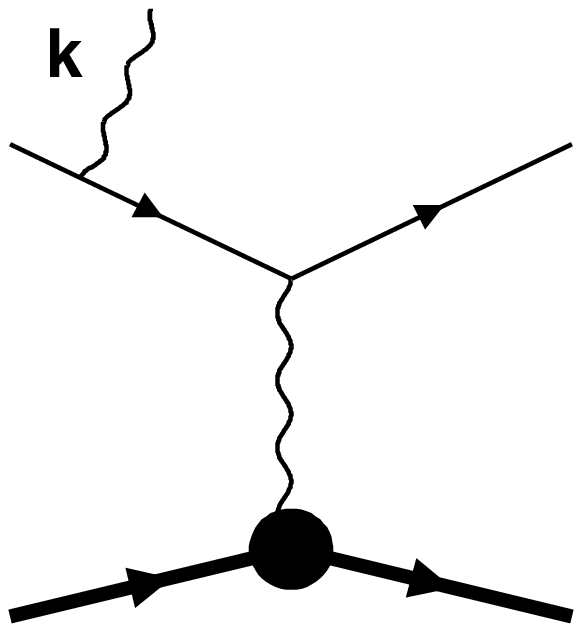}
\put(-36,36){\mbox{(d)}}
}
\end{picture}
&
\begin{picture}(20,20)
\put(40,0){
\epsfxsize=6cm
\epsfysize=10cm
\epsfbox{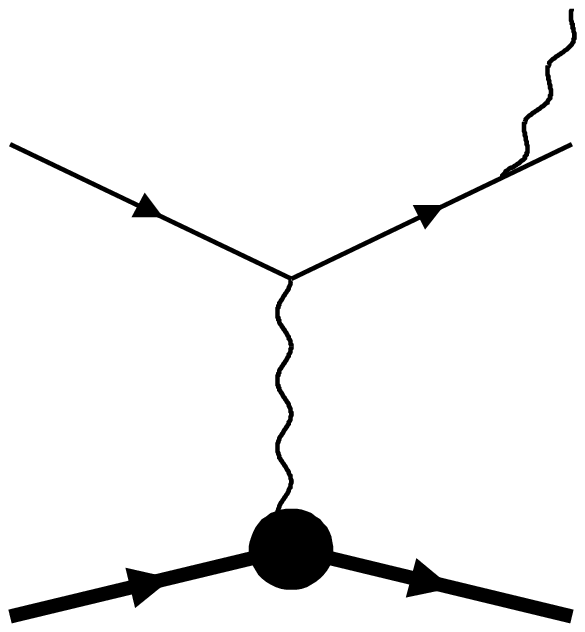}
\put(-36,36){\mbox{(e)}}
}
\end{picture}
\end{tabular}
\vspace{-35mm}
\caption{\protect\it
Feynman graphs contributing to the non-radiated (a, b, c) and
the radiated (d, e)
parts of the cross sections in lepton-nucleus scattering.}
\end{figure}
\begin{equation}
\sigma ^{non-rad}(v_{min}) =
\sigma _{0}e^{\delta_{inf}}
(1+ \delta_{VR}+\delta_{vac}
+\delta^{add}(v_{min}))
+\sigma^{add}_{R}(v_{min}).
\end{equation}
The explicit expressions for $\sigma _{0}$, $\delta_{inf}$,
$\delta_{VR}$ can be found in \cite{AAM}. The dependence of the
non-radiative part of the cross section on $v_{min}$ appears in
the following way:
\begin{equation}
\delta ^{add}(v_{min})=
\frac{2\alpha }{\pi}
\log \left (\frac {v_{max}} {v_{min}} \right )
\left [1-\log \left (\frac{Q^2}{m^2} \right)\right],
\end{equation}
\begin{equation}
\sigma _R^{add}{(v_{min})}
=-{\alpha^3 \over 2S^2}\int\limits_{t'_1}^{t'_2}
dt \sum_i \left [
\theta_{i}
{{\mathcal F}_i(t) \over t^2}-
4 \theta^B_{i}
{\mathcal F}_{IR}
{{\mathcal F}_i(Q^2) \over Q^4}
\right ].
\label{t}
\end{equation}
Here $S=2k_1p_1$, $v_{max}=S-Q^2-M^2Q^2/S$,
${\mathcal F}_i(t)$ are the squared combinations of the electric
and
magnetic elastic form factors. The quantities $\theta _{i}$ as well as
$\theta ^B _i$ are the analytical functions \cite{AAM} of
kinematic
invariants for the radiative and Born subprocesses, respectively.

\begin{figure}[t]
\unitlength 1mm
\vspace*{5mm}
\begin{picture}(160,80)
\put(30,0){
\epsfxsize=10cm
\epsfysize=8cm
\epsfbox{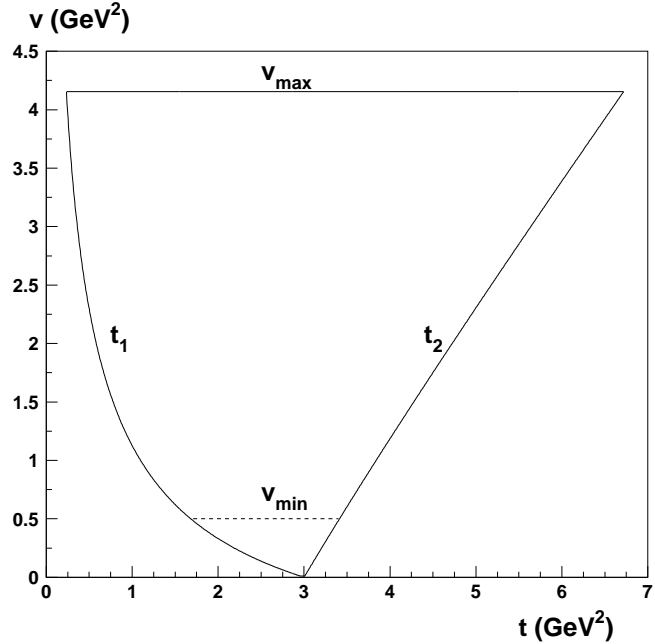}
}
\end{picture}
\vspace*{-5mm}
\caption{\label{kinreg}
The region of integration over $v$- and $t$-variables
for JLab kinematics ($Q^2=3$ GeV$^2$, $S=7.5$ GeV$^2$).
}
\end{figure}
The radiative part of the total cross section can be presented as
an integral over the real photonic phase space in the following way:
\begin{equation}
\sigma ^{rad}(v_{min})
= -{\alpha^3 \over 2S^2}
\int\limits_{t_1}^{t_2}dt
\sum_i
{{\mathcal F}_i(t) \over t^2}
\Theta_{i}(t),
\label{tt}
\end{equation}
where
\begin{equation}
\Theta_{i}(t)
=
\int\limits_{v_{1}}^{v_{max}}dv
\theta_{i}
(t, v)
=
\int\limits_{v_{1}}^{v_{max}}dv
\int\limits_{0}^{2\pi}d\phi _k
\theta_{i}
(t, v,\phi _k),
\label{vp}
\end{equation}
and
\begin{eqnarray}
v_1&=&max\{ v_{min},
\frac{(t-Q^2)(\sqrt{t}\mp\sqrt{4M^2+t})}{2\sqrt{t}} \},\;\;
\\
t_{1,2}
&=&\frac{2M^2Q^2+v_{max}(Q^2
+v_{max}\mp
\sqrt{(Q^2+v_{max})^2+4M^2Q^2})}{2(M^2+v_{max})},
\\
t'_{1,2}&=&t_{1,2}(v_{max} \rightarrow v_{min}).
\end{eqnarray}
The radiative photonic phase space $t$- and $v$-variables are
presented in Figure \ref{kinreg}.

Notice that in (\ref{t}) as well as in (\ref{tt}) the structure
functions ${\mathcal F}_i$ depend on only one
integration
variable, namely, on $t$.  Therefore to apply our
program to different fits or  models of nucleon form factors
in (\ref{t}) and (\ref{tt}), the numerical integration over the variable
$t$ is used. On the other hand, to speed up the process of event
generation, integration of the expressions  $\theta _i$
in (\ref{vp}) over the other photonic variables $v$ and $\phi _k$
has been performed
analytically because the quantity $\theta_{i}(t, v,\phi _k)$
appears as a result of the calculation of the matrix element squares
with the real photon emission and do not depend on any model for
elastic electron--nucleon scattering.

\begin{figure}[!t]
\vspace{14mm}
\unitlength 1mm
\begin{tabular}{cc}
\begin{picture}(60,60)
\put(-6,0){
\epsfxsize=9cm
\epsfysize=8cm
\epsfbox{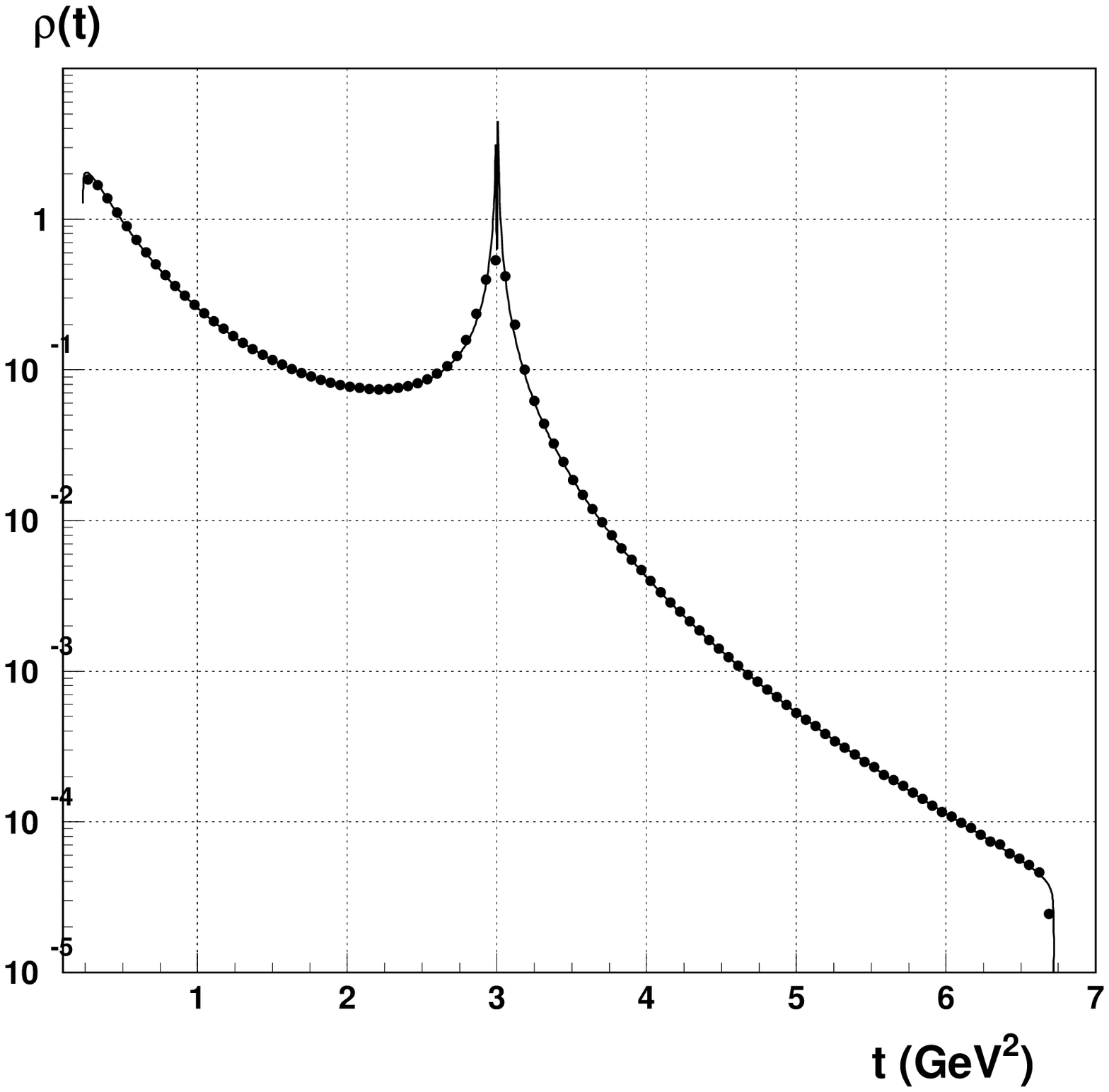}
}
\end{picture}
&
\begin{picture}(60,60)
\put(12,0){
\epsfxsize=9cm
\epsfysize=8cm
\epsfbox{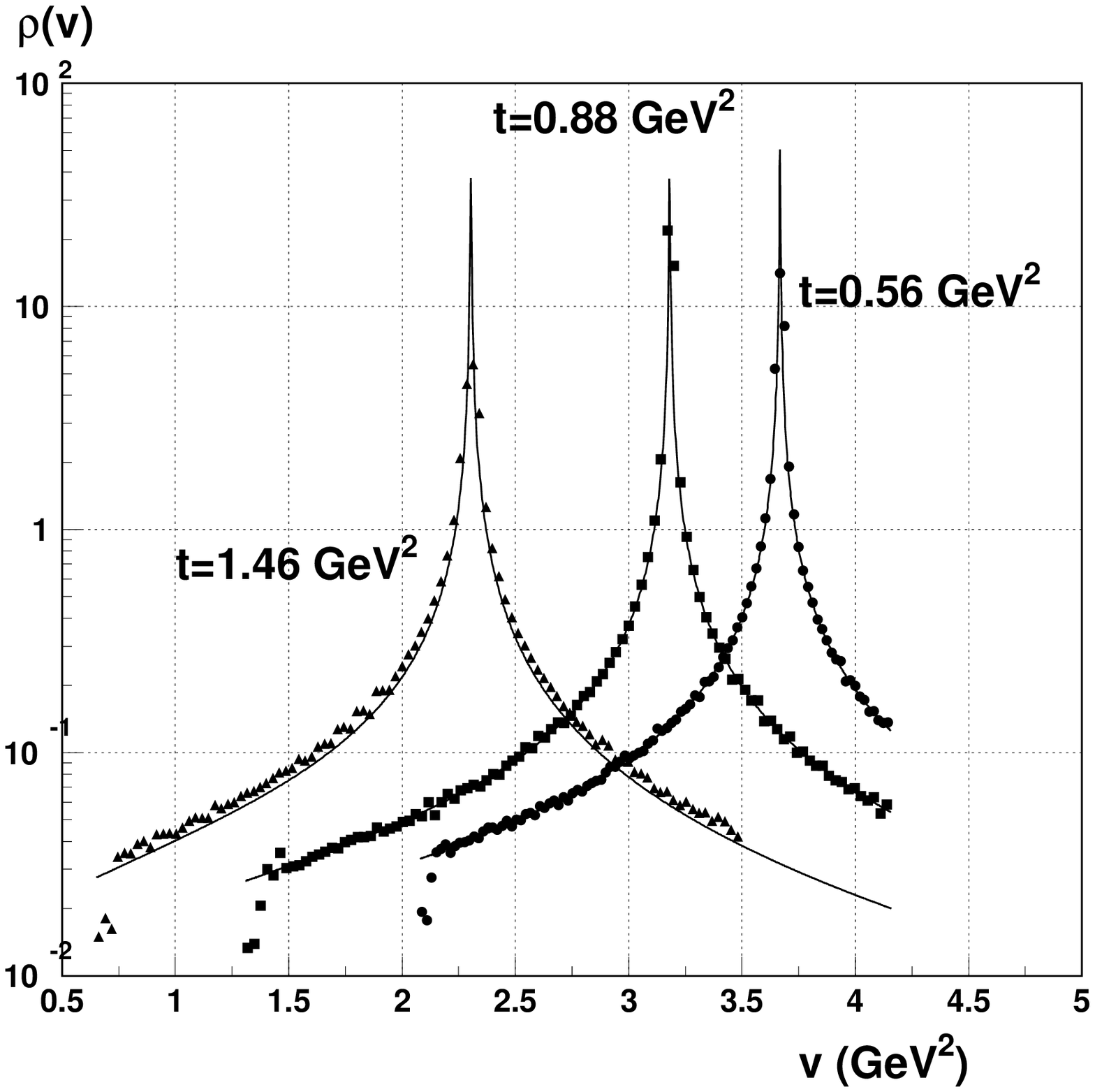}
}
\end{picture}
\end{tabular}
\label{fig2}
\vspace{-5mm}
\caption{
$t$- and $v$-histograms and the corresponding probability
densities for the kinematics $Q^2=3$ GeV$^2$ and $S=7.5$ GeV$^2$.
}
\end{figure}

\section{Simulation of radiative events}

To simulate $v$ and $\phi _k$ in accordance with the
analytically calculated distributions without any numerical
integration over $t$, the first step is simulation of the variable $t$.
Then taking into account the simulated
value $t$, the distribution function for $v$ can be constructed.
Finally, $\phi _k$
is simulated according to the simulated values of $t$ and $v$.
If the event generation is performed successfully, the event
distributions over these three variables has to
follow the corresponding probability densities:
\begin{equation}
\begin{array}{ll}
\displaystyle
\rho (t)= \frac 1{N_t}\sum _i
{{\mathcal F}_i(t)}
\Theta_{i} (t)/t^2,&
\displaystyle
N_t=
\int\limits_{t_1}^{t_2}dt
\sum_i
{{\mathcal F}_i(t)}
\Theta_{i} (t)/t^2
\\
\displaystyle
\rho (v)=\frac 1{N_v}
\sum \limits_i{\mathcal F}_i(t_g)\theta_{i}(t_g, v),&
\displaystyle
N_v=\sum \limits_i{\mathcal F}_i(t_g)\Theta_{i}(t_g)
\\
\displaystyle
\rho (\phi_k)=\frac 1{N_{\phi _k}}
\sum \limits_i{\mathcal F}_i(t_g)\theta_i
(t _g, v _g,\phi _k),&
\displaystyle
N_{\phi _k}=\sum \limits_i{\mathcal F}_i(t_g)
\theta_{i}(t_g, v_g)
\end{array}
\end{equation}

As an example, the simulated distributions of the kinematic
variables $t$ and $v$ together with their probability densities
for representative kinematics at CEBAF energies are compared in Figure 3.

In summary, we presented the first results of the FORTRAN code
ELRADGEN, which is a Monte Carlo generator of elastic unpolarized
$ep$--scattering that includes electromagnetic radiative events.

\end{document}